\documentclass{article}
\usepackage{slashed,feynmf,epsfig,amsmath,amssymb,enumitem}
\usepackage[papersize={8.5in,11in},margin=1in]{geometry}

\usepackage{graphicx}
\usepackage{hyperref}
\newcommand{\be}{\begin{equation}}
\newcommand{\ee}{\end{equation}}
\def\bea{\begin{eqnarray}}  \def\eea{\end{eqnarray}}

\usepackage{xcolor}
\definecolor{MyRed}{HTML}{BE0032}

\begin{document}
\title{Finite Quantum Field Theory and Renormalization Group}
\author{M. A. Green$^a$ and J. W. Moffat$^{a,b}$\\~\\
  \emph{\small $^a$Perimeter Institute for Theoretical Physics,
    Waterloo, Ontario
    N2L 2Y5, Canada}\\
  \emph{\small $^b$Department of Physics and Astronomy, University of
    Waterloo, Waterloo, Ontario N2L 3G1, Canada}}
\maketitle

%\date{\today}

%\thanks{PACS: 98.80.C; 04.20.G; 04.40.-b}

% ----------------------------------------------------------------

\begin{abstract}
  \noindent Renormalization group methods are applied to a scalar
  field within a finite, nonlocal quantum field theory formulated
  perturbatively in Euclidean momentum space. It is demonstrated that
  the triviality problem in scalar field theory, the Higgs boson mass
  hierarchy problem and the stability of the vacuum do not arise as
  issues in the theory. The scalar Higgs field has no Landau pole.
\end{abstract}

\maketitle

\section{Introduction}
An alternative version of the Standard Model (SM), constructed using
an ultraviolet finite quantum field theory with nonlocal field
operators, was investigated in previous work~\cite{Moffat2018}. In
place of Dirac delta-functions, $\delta(x)$, the theory uses
distributions ${\mathcal E}(x)$ based on finite width Gaussians. The
Poincar\'e and gauge invariant model adapts perturbative quantum field
theory (QFT), with a finite renormalization, to yield finite quantum
loops.  For the weak interactions, $SU(2)\times U(1)$ is treated as an
{\it ab initio} broken symmetry group with non-zero masses for the $W$
and $Z$ intermediate vector bosons and for left and right quarks and
leptons.  The model guarantees the stability of the vacuum.  Two
energy scales, $\Lambda_M$ and $\Lambda_H$, were introduced; the rate
of asymptotic vanishing of all coupling strengths at vertices not
involving the Higgs boson is controlled by $\Lambda_M$, while
$\Lambda_H$ controls the vanishing of couplings to the Higgs.
Experimental tests of the model, using future linear or circular
colliders, were proposed.  Present observations are consistent with
\mbox{$\Lambda_M \geq 10$ TeV}. The Higgs boson mass hierarchy
problem will be solved if future experiments confirm the prediction
$\Lambda_H\lesssim 1$ TeV.

In the following, we will investigate the consequences of an
application of renormalization group (RG) methods for the perturbative
finite renormalizable model. We will concentrate on a nonlocal spin 0
scalar field $\phi=\phi_H$ Lagrangian model which is perturbatively
formulated in Euclidean momentum space and might describe the Higgs
boson field if nonlocality were fundamental. The ultraviolet finite
theory resolves the Higgs mass hierarchy problem, the scalar field
model triviality problem and removes the Landau pole singularity for
the Higgs field.

\section{Scalar Field Theory}

The Lagrangian we consider for a real scalar field $\phi\equiv\phi_H$
describing the Higgs boson in Euclidean space is%
\be%
{\cal L}_H=\frac{1}{2}(-\phi\Box\phi+m_0^2\phi^2)+\frac{1}{4!}\lambda_0\phi^4.%
\ee%

Using the formalism of \cite{KleppeWoodard1993}, we assume that the
vacuum expectation of the bare field $\phi$ vanishes and write
$\phi=Z^{1/2}\phi_r$, where $\phi_r$ is the renormalized field.
Expressed as series expansions in powers of the physical coupling
$\lambda$, mass $m$ and energy scale $\Lambda_H$, the field strength
renormalization constant $Z$ and the bare parameters $m_0$ and
$\lambda_0$ are given by:%
\be%
Z=1+\delta Z(\lambda, m, \Lambda_H^2),%
\ee%
\be%
Zm_0^2=m^2+\delta m^2(\lambda, m,\Lambda_H^2),%
\ee%
\be%
Z^2\lambda_0=\lambda+\delta \lambda(\lambda, m, \Lambda_H^2).%
\ee%

The propagator in Euclidean momentum space is given by
\be%
i\Delta_H(p)\equiv \frac{i{\mathcal E^2(p)}}{p^2+m^2},
\ee%
where ${\cal E}(p)$ is the entire function:
\be%
{\mathcal E}(p)=\exp\biggl[-\biggl(\frac{p^2+m^2}{2\Lambda_H^2}\biggr)\biggr].%
\ee%

Evaluating the one-loop self-energy graph gives a constant shift to
the Higgs boson bare self-energy~\cite{KleppeWoodard1993}:%
\be%
\label{OneLoop}
-i\Sigma_0=\frac{-iZ^{-2}\lambda}{32\pi^2}m^2\,\Gamma\biggl(-1,\frac{m^2}{\Lambda_H^2}\biggr),%
\ee%
where $\Gamma(n,z)$ is the incomplete gamma function:%
\be%
\label{Gamma}
\Gamma(n,z)=\int_z^\infty
dt\,t^{n-1}\exp(-t)=(n-1)\Gamma(n-1,z)+z^{n-1}\exp(-z).%
\ee%
Setting $n=0$ in (\ref{Gamma}) gives:%
\be%
\Gamma(0,z)=E_1(z)=\int_z^\infty dt\frac{\exp(-t)}{t} =
-\ln(z)-\gamma-\sum^{\infty}_{n=1}\frac{(-z)^n}{nn!},%
\ee%
%and%
\be%
\Gamma(-1,z)=-\Gamma(0,z)+\frac{\exp(-z)}{z}.%
\ee%

The renormalized one-loop self-energy $\Sigma_R(p^2)$ can then be
written in the form:%
\be%
\Sigma_R(p^2)=\delta Z(p^2+m^2)+\delta m^2
+\frac{Z^{-1}\lambda}{32\pi^2}m^2\Gamma
\biggl(-1,\frac{m^2}{\Lambda_H^2}\biggr)+{\cal O}(\lambda^2).%
\ee%
The renormalized mass and field strength are given by
\be%
\delta m^2=-\frac{\lambda}{32\pi^2}m^2\Gamma\biggl(-1,\frac{m^2}{\Lambda_H}\biggr)+{\cal O}(\lambda^2),
\ee%
\be%
\delta Z={\cal O}(\lambda^2).%
\ee%
The expansion of the one-loop Higgs boson self-energy mass correction for $m \ll \Lambda_H$ is%
\be%
\label{masscorrection}
\delta m^2=\frac{\lambda}{32\pi^2}
\biggl[-\Lambda_H^2+m^2\ln\biggl(\frac{\Lambda_H^2}{m^2}\biggr)
+m^2(1-\gamma)+{\cal O}\biggl(\frac{m^2}{\Lambda_H^2}\biggr)\biggr]+{\cal O}(\lambda^2).%
\ee%
The one-loop vertex correction is given by%
\be%
\label{vertexcorrection}
\delta\lambda=\frac{3\lambda^2}{16\pi^2}\int_0^{1/2}dx\,
\Gamma\biggl(0,\frac{1}{1-x}\frac{m^2}{\Lambda_H^2}\biggr) +{\cal O}(\lambda^3).%
\ee%
For $m \ll \Lambda_H$ this can be expanded for the Higgs boson to give%
\be%
\label{expandvertex}
\delta\lambda=\frac{3\lambda^2}{16\pi^2}
\biggl[\frac{1}{2}\ln\biggl(\frac{\Lambda_H^2}{m^2}\biggr)
+\frac{1}{2}(\ln(2)-1-\gamma)+{\cal O}
\biggl(\frac{m^2}{\Lambda_H^2}\biggr)\biggr]+{\cal O}(\lambda^3).%
\ee%

\section{Callan-Symanzik Equation and Running of $\lambda$}

Let us consider the Callan-Symanzik equations~\cite{Callan1970,Symanzik1970,Symanzik1971,Peskin1995} satisfied with our energy (length) scales $\Lambda_i$ playing the roles of finite renormalization scales. In finite QFT theory, the equations for the regularized amplitudes $\Gamma^{(n)}(x-x')$ are
\be
\biggl[\Lambda_i\frac{\partial}{\partial\Lambda_i}+\beta(g_i)\frac{\partial}{\partial g_i}-2\gamma(g_i)\biggr]\Gamma^{(n)}=0,
\ee
where $g_i$ are the running coupling constants associated with diagram vertices. The correlation functions will satisfy this equation for the n-th order $\Gamma^{(n)}$ for the Gell-Mann-Low functions $\beta(g_i)$ and the anomalous dimensions in nth-loop order.

For the Higgs field, the RG equation is given by%
\be
\biggl[\Lambda_H\frac{\partial}{\partial\Lambda_H}+\beta(\lambda)\frac{\partial}{\partial\lambda}-2\gamma(\lambda)\biggr]\Gamma^H=0.
\ee%
where the coupling $\lambda$ runs with $\Lambda_H$.  Neglecting the
anomalous dimension term $\gamma(\lambda)$ and replacing the measured
Higgs mass $m$ by the RG scaling mass $\mu$ yields the equation:%
\be
\label{RGequation}
\beta(\lambda)=
-\frac{d\lambda}{d\ln\left(\frac{\Lambda_H}{\mu}\right)}.
\ee%

We obtain from (\ref{vertexcorrection}) the Higgs field $\beta$ function:
\be
\label{betafunction}
\beta(\lambda)=\frac{3\lambda^2}{16\pi^2}I(\mu^2/\Lambda_H^2)+{\cal O}(\lambda^3),
\ee%
where
\be
I(\mu^2/\Lambda_H^2)=\int_0^{1/2}dx\,\Gamma\biggl(0,\frac{1}{1-x}\frac{\mu^2}{\Lambda^2_H}\biggr).
\ee
Using the identities $\Gamma(0,y)=E_1(y)=-{\rm Ei}(-y)$, yields:%
\bea%
\nonumber I(\mu^2/\Lambda_H^2)=-\int_0^{1/2}dx\, {\rm
  Ei}\biggl(-\frac{1}{1-x} \frac{\mu^2}{\Lambda^2_H}\biggr)=&\!\!
\frac{1}{2}\!\left(\exp\!\left(\frac{-2 \mu^2}{\Lambda_H^2}\right) +
  \left(1 + \frac{2 \mu^2}{\Lambda_H^2}\right) {\rm
    Ei}\!\left(\frac{-2 \mu^2}{\Lambda_H^2}\right)\right)\\
~&\hspace*{-.3in}-\exp\!\left(\frac{-\mu^2}{\Lambda_H^2}\right) -
\left(1 + \frac{\mu^2}{\Lambda_H^2}\right){\rm
  Ei}\!\left(\frac{-\mu^2}{\Lambda_H^2}\right).%
\eea%
We have $\lambda=\lambda_0+\delta\lambda$ and
\be
\frac{d\lambda}{d\left(\frac{\Lambda_H}{\mu}\right)}=\frac{d\delta\lambda}{d\ln\left(\frac{\Lambda_H}{\mu}\right)}=-\beta(\lambda).
\ee
From (\ref{betafunction}) we obtain
\be
\frac{d\lambda}{\lambda^2}=-\frac{3}{16\pi^2}dI(\mu^2/\Lambda_H^2).
\ee
Integrating this equation we get
\be
\label{inverselambdaren}
\frac{1}{\lambda}=\frac{1}{\lambda_0}+J(\mu^2/\Lambda_H^2),
\ee
where
\be
J(\mu^2/\Lambda_H^2)=\frac{3}{16\pi^2}\int\frac{d\Lambda_H}{\Lambda_H}I(\mu^2/\Lambda_H^2).
\ee

Evaluating the integral for $J(\mu^2/\Lambda_H^2)$, using $x=\frac{\mu^2}{\Lambda_H^2}$, gives%
\bea%
\nonumber J(x)&=&\frac{3}{128\pi^2}\left(-2\exp(-2 x) + 4\exp(-x) +
  \pi^2 - (2+4
  x){\rm Ei}(-2 x) + (4+4 x){\rm Ei}(-x)\hspace*{\fill}\right.\\
\nonumber &~&+4 x \,_3{\rm F}_3(1,1,1;2,2,2;-2 x) - 4 x \,_3{\rm
    F}_3(1,1,1;2,2,2;-x) \\
&~&- \left. \ln(2)^2 - \ln(4)\gamma + 2(
  \gamma-\ln(2))\ln(x) + \ln(x)^2\right),%
\eea%
where $_p{\rm F}_q(a_1,...,a_p;b_1,...,b_q;z)$ is a generalized
hypergeometric function.

From (\ref{inverselambdaren}) we obtain:
\be
\lambda=\frac{\lambda_0}{1+\lambda_0J(\mu^2/\Lambda_H^2)},
\ee
or
\be
\lambda_0=\frac{\lambda}{1-\lambda J(\mu^2/\Lambda_H^2)}.
\ee
We can compare (\ref{inverselambdaren}) with the equation obtained in SM:
\be
\label{SMresult}
\frac{1}{\lambda}=\frac{1}{\lambda_0}+\frac{3}{16\pi^2}\ln\biggl(\frac{\Lambda_C}{\mu}\biggr),
\ee
or
\be
\lambda=\frac{\lambda_0}{1+\frac{3\lambda_0}{16\pi^2}\ln\left(\frac{\Lambda_C}{\mu}\right)},
\ee
and
\be
\lambda_0=\frac{\lambda}{1-\frac{3\lambda}{16\pi^2}\ln\left(\frac{\Lambda_C}{\mu}\right)}.
\ee
In the SM, the $\lambda\phi^4$ model is renormalizable and produces finite scattering amplitudes and cross sections, but renormalization theory demands that the cutoff $\Lambda_C$ must be taken to infinity, $\Lambda_C\rightarrow\infty$~\cite{Dyson1949,Dyson1952}. Then, from (\ref{SMresult}), the renormalized coupling constant $\lambda=0$. This is known as the triviality problem~\cite{Landau1955,LandauPomeranchuk1955,Landau1956,Callaway1986,Callaway1988,Wilson}. This result holds even in the limit $\lambda_0\rightarrow\infty$:
\be
\frac{1}{\lambda}\sim\frac{3}{16\pi^2}\ln\biggl(\frac{\Lambda_C}{\mu}\biggr).
\ee

In the earlier paper~\cite{Moffat1990}, it was demonstrated that the triviality problem for the scalar field field could be resolved in the finite QFT theory. Because $\Lambda_H=1/{\ell_H}$ is a fundamental constant to be measured, it cannot be taken to infinity as in the case of infinite renormalization theory. Thus, we cannot take the limit $\ell_H\rightarrow0$ corresponding to the $\delta$-function limit. From Fig. 1, we observe that when we choose $\Lambda_H\lesssim 1$ TeV, the Higgs mass hierarchy problem is resolved, for we have $\delta m^2/m^2\sim {\cal O}(1)$ where $m=125$ GeV. From Fig. 1, we observe that for $\Lambda_H > \frac{1}{2}\mu$, we avoid a Landau pole and, in particular, for $700 < \Lambda_H < 1$ TeV, we resolve the triviality problem for the scalar Higgs field and the Higgs mass fine-tuning hierarchy problem.

\begin{figure}[b!]
  \center\includegraphics[height=3.75in]{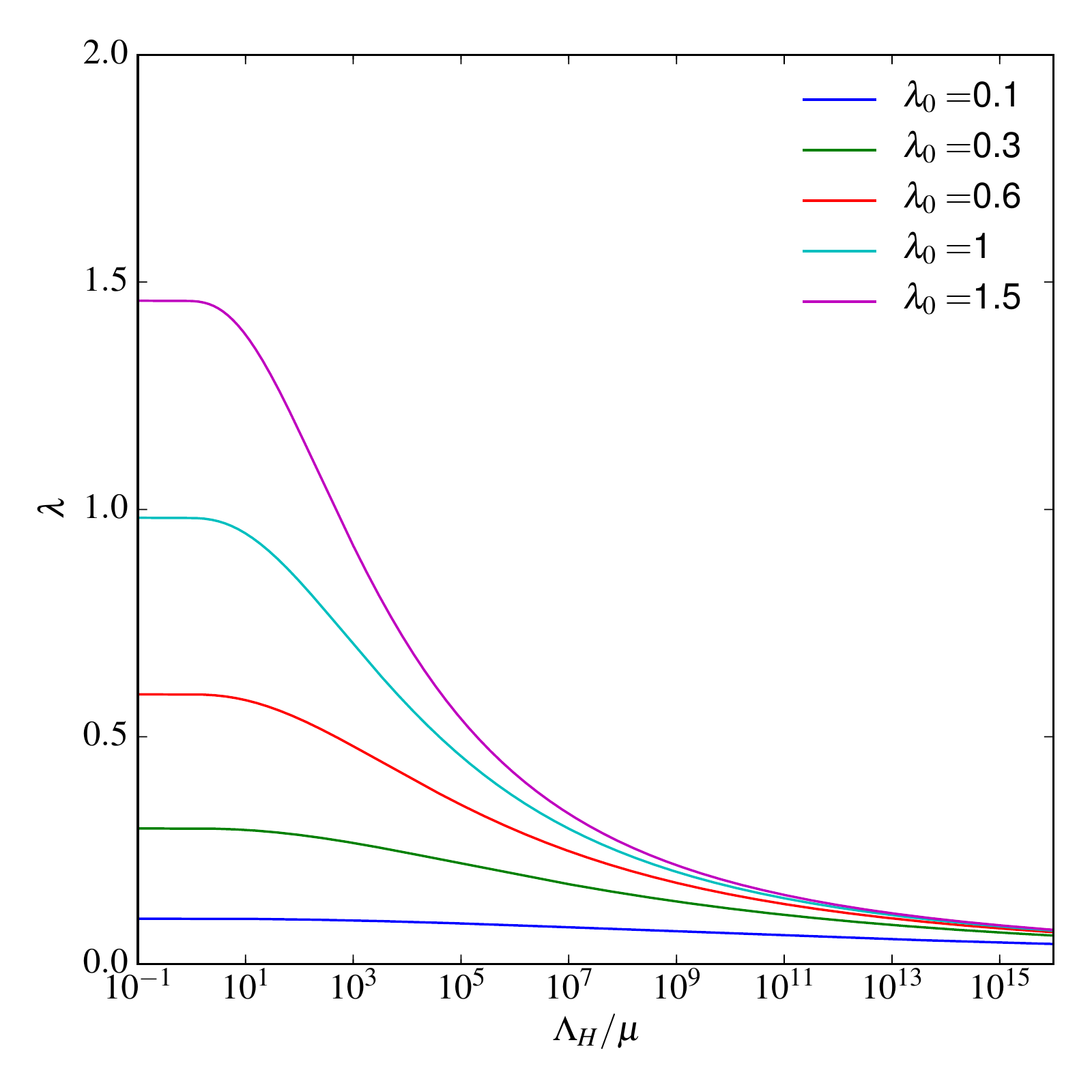}
  \caption{Running of $\lambda$ versus $\Lambda_H/\mu$ for
    finite QFT.}
\label{F:ASD}
\end{figure}

Choosing an energy $\mu_0$ above $\Lambda_H$ as a measurement probe of
the running of $\lambda$ is attempting to make a measurement within
the finite Gaussian distribution length size
$\ell_H$~\cite{Moffat2018} and is prohibited within the perturbation
approximations we have assumed. The results obtained for the running
of $\lambda$ are for a single Higgs particle interacting with another
Higgs particle. This cannot describe a fully realistic situation, for
the Higgs coupling to other particles such as the top quark (the top
quark-Higgs coupling $\lambda_t\sim {\cal O}(1)$) may play an
important role.

The above one-loop calculations have employed a perturbative
formulation in Euclidean momentum space and rely on analytic
continuation to obtain corresponding Lorentzian results. At tree
level, the theory is completely equivalent to the classical field
theory with the same Lagrangian. Although loop diagrams should be
finite at all levels, convergence of the quantum perturbative
formalism has not been demonstrated. Because the Fourier transform to
momentum space is generally not well defined in curved spacetime, no
claims can be made about whether or how the theory might be applied in
the context of an expanding universe; the energy scales $\Lambda_M$,
$\Lambda_H$ may well depend on emergent and evolving properties of the
classical universe (e.g., entropy density), thus bridging the gap
between quantum and classical. Whether a nonperturbative quantum
formulation can be developed remains to be determined.

\section*{Acknowledgments}

Research at the Perimeter Institute for Theoretical Physics is supported by the Government of Canada through industry Canada and by the Province of Ontario through the Ministry of Research and Innovation (MRI).

\end{document}